\documentclass[seceq]{ptptex}
\usepackage{wrapft}
\usepackage{graphicx}

\newcommand{\ket}[1]{| {#1} \rangle}     
\newcommand{\wtilde}[1]{\widetilde{#1}} 

\def\bsub{\begin{subequations}}
\def\esub{\end{subequations}}
\def\beq{\begin{eqnarray}}
\def\eeq{\end{eqnarray}}
\def\bsub{\begin{subequations}}
\def\esub{\end{subequations}}
\def\b{\begin{equation}}
\def\bs{\begin{split}}
\def\es{\end{split}}
\def\e{\end{equation}}

\begin{document}

\title{Spontaneous magnetization in high-density quark matter}

\author{Yasuhiko {\sc Tsue}$^{1,2}$, {Jo\~ao da {\sc Provid\^encia}}$^{1}$, {Constan\c{c}a {\sc Provid\^encia}}$^{1}$,\\
{Masatoshi {\sc Yamamura}}$^{1,3}$ and {Henrik {\sc Bohr}}$^{4}$
}

\inst{$^{1}${CFisUC, Departamento de F\'{i}sica, Universidade de Coimbra, 3004-516 Coimbra, 
Portugal}\\
$^{2}${Physics Division, Faculty of Science, Kochi University, Kochi 780-8520, Japan}\\
$^{3}${Department of Pure and Applied Physics, 
Faculty of Engineering Science, Kansai University, Suita 564-8680, Japan}\\
$^{4}${Department of Physics, B.307, Danish Technical University, DK-2800 Lyngby, Denmark}
}

\abst{
It is shown that the spontaneous magnetization occurs due to the anomalous magnetic moments of 
quarks in the high-density quark matter under the tensor-type four-point interaction. 
The spin polarized condensate for each flavor of quark appears at high baryon density, which 
leads to the spontaneous magnetization due to the anomalous magnetic moments of quarks. 
The implications to the strong magnetic field in the compact stars is discussed. 
}


\maketitle

\maketitle

\section{Introduction}

One of the recent interests about the physics governed by quantum chromodynamics (QCD) 
may be to clarify the phase structure of QCD 
with respect to the temperature, baryon chemical potential 
or baryon number density, external magnetic field and so on \cite{FH}. 
In the region with high temperature and low density, the lattice QCD simulation gives many 
insights about the QCD world.  
However, as is well known, in the region of large quark chemical potential, 
there are some problems to be solved 
in order to calculate the physical quantities definitely.   
But, it is expected that there are still many interesting phenomena in the region of low temperature 
and large chemical potential. Especially, it is expected that 
there are many exotic phases such as the two-flavor color superconducting phase, 
the color-flavor locked phase \cite{ARW,IB,CFL}, the quarkyonic phase \cite{MaC}, the phase with the inhomogeneous condensate 
\cite{NT} and so forth.

In the heavy-ion collision experiments such as relativistic heavy-ion collider  (RHIC) 
experiments at Brookhaven National Laboratory, 
it is believed that the quark-gluon phase is realized apart from the 
hadronic phase. 
Thus, the quark matter may be created and 
the extreme states of QCD with finite temperature and density may be realized 
in the heavy-ion collision experiments.
On the other hand, the high-density hadronic phase or quark phase may be realized 
in the inner core of the compact star objects such as the neutron stars, magnetars and 
quark stars, if they exist. 
In the core of these compact stars, it is expected that the hadron or quark phase exists 
at low temperature and high baryon density. 
For example, the quark matter is considered to exist 
in the core of pulsars with two solar masses in compact stars, and in the case of 
the heavy ion collisions \cite{Blaschke}. 
Therefore, the investigation of quark matter in the region of the low temperature and 
the large quark chemical potential is one of very interesting and important subjects in 
order to understand the whole world governed by QCD. 

It is further 
known that neutron stars, especially the so-called magnetars \cite{magnetar1,magnetar2}, have a strong magnetic field. 
However, the origin of the strong magnetic field is not so clear. 
It has been pointed out that the strong magnetic field may be created 
if the quark liquid exists in the core of the compact stars \cite{Tatsumi}.  
Further, the possibility of the quark spin polarization in the high-density quark matter has been investigated when the 
pseudovector-type interaction between quarks exists \cite{NMT}. 
Under the pseudovector-type interaction, it was shown that the spin polarized condensate appears, 
which leads to a ferromagnetism in a quark matter \cite{TMN}. 
However, the spin polarized condensate only appears in a narrow region of the quark chemical potential.

As for the spin polarized condensate, it has been shown that the tensor-type interaction between quarks leads to 
the spin polarized phase in high-density quark matter \cite{BJ,arXiv,oursPTP}. 
The present authors have shown the possibility of the quark spin polarized phase in the quark matter at high baryon density 
against the two-flavor color superconducting phase \cite{oursPTEP1} and the color-flavor locked phase \cite{oursPTEP2} 
under both the quark pairing interaction and tensor-type four-point interaction in the Nambu-Jona-Lasinio (NJL)-type model \cite{NJL,Buballa}. 
However, the magnetic feature is not considered in the tensor-type interaction until now .

In this paper, the magnetic features are investigated under the existence of the spin polarized condensate in the
NJL model with the tensor-type four-point interactions. 
It will be shown that the anomalous magnetic moments of quarks \cite{AMM} play an essential role, the existence of which leads to the 
spontaneous magnetization of quark matter. 
Also, the implication to the magnetic field of compact stars such as neutron stars with quark matter, namely hybrid stars, 
will be discussed. 

This paper is organized as follows: In the next two sections, \S 2 and \S 3, the thermodynamic potential 
under the external magnetic field is given with and without the spin polarized condensate. 
In section \S 4, an approximate expression of the thermodynamic potential with small magnetic field is derived. 
In section \S 5, it is shown that the spontaneous magnetization does not appear in the case of no anomalous magnetic moments of quarks. 
In section \S 6, the anomalous magnetic moments of quarks are introduced within the mean field approximation. 
As a result, spontaneous magnetization occurs in quark matter in the region of high baryon density 
due to both the spontaneous spin polarization and the anomalous magnetic moments of quarks. 
In section \S 7, the implication to the hybrid compact stars is discussed briefly and it is shown that 
the strong magnetic field in the surface of the compact stars can be revealed. 
The last section is devoted to a summary and concluding remarks.

\setcounter{equation}{0} 
\section{Thermodynamic potential under external magnetic field}

We consider the two-flavor case. 
Let us start from the following Lagrangian density with chiral symmetry:
\beq\label{2-1}
{\cal L}_0={\bar \psi}i\gamma^{\mu}\partial_{\mu}\psi-\frac{G}{4}(
({\bar \psi}\gamma^{\mu}\gamma^{\nu}{\vec \tau}\psi)({\bar \psi}\gamma_{\mu}\gamma_{\nu}{\vec \tau}\psi)
+({\bar \psi}i\gamma_5\gamma^{\mu}\gamma^{\nu}\psi)({\bar \psi}i\gamma_5\gamma_{\mu}\gamma_{\nu}\psi))\ , 
\eeq
where ${\vec \tau}$ represents the flavor $su(2)$-generator.

In previous study, we have found that a spin polarized condensate\break 
$F_3=-G\langle{\bar \psi}\Sigma_3\tau_3\psi\rangle$ 
may be realized at high baryon density, where $\Sigma_3=-i\gamma^1\gamma^2$ is the spin operator. 
Thus, the Lagrangian density under the mean field approximation is obtained as 
\beq\label{2-2}
& &{\cal L}_{\rm MF}=i{\bar \psi}\gamma^{\mu}\partial_{\mu}\psi-F_3{\bar \psi}\Sigma_3\tau_3\psi-\frac{F^2}{2G}\ , 
\\
& &\ \ F_3=-G\langle{\bar \psi}\Sigma_3\tau_3\psi\rangle=F\tau_f\ , \qquad
\Sigma_3=-i\gamma^1\gamma^2=\left(
\begin{array}{cc}
\sigma_3 & 0 \\
0 & \sigma_3
\end{array}
\right)\ , 
\nonumber
\eeq 
where $\tau_f=1$ for $u$-quark $(f=u)$ and $\tau_f=-1$ for $d$-quark ($f=d$), denote the eigenvalues of $\tau_3$. 
Here, $\sigma_3$ is the third component of the Pauli spin matrices. 

Hereafter, let us consider the system under the external magnetic field $B$ along the $z$-axis. 
The Lagrangian density is recast into
\beq\label{2-3}
{\cal L}=i{\bar \psi}\gamma^{\mu}D_{\mu}\psi-F_3{\bar \psi}\Sigma_3\tau_3\psi-\frac{F^2}{2G}\ , 
\eeq
where $D_{\mu}$ represents the covariant derivative: 
\beq\label{2-4}
D_{\mu}=\partial_{\mu}+iQA_{\mu}\ , \qquad
A_{\mu}=\left(0,\ \frac{By}{2},\ -\frac{Bx}{2},\ 0\right)=\left(0,\ -{\mib A}\right)\ . 
\eeq
Here, for up (down) quark, $Q=2e/3$ ($-e/3$) where $e$ is the elementary charge. 
Because we investigate the quark matter at finite density system, the Hamiltonian density 
with quark chemical potential $\mu$ is obtained as 
\beq\label{2-5}
{\cal H}-\mu{\cal N}&=&{\bar \psi}(-i{\mib \gamma}\cdot
({\mib \nabla}-iQ{\mib A})-\mu\gamma^0)\psi +F_3{\bar \psi}\Sigma_3\tau_3\psi+\frac{F^2}{2G} \nonumber\\
&=&\psi^{\dagger}(h-\mu)\psi+\frac{F^2}{2G}\ ,
\eeq
where ${\cal N}$ represents the quark-number operator. 
Here, $h$ can be expressed as 
\beq\label{2-6}
h&=&{\mib \alpha}\cdot({\hat {\mib p}}-Q{\mib A})+F_3\tau_3\beta\Sigma_3\nonumber\\
&=&\alpha_z{\hat p}_z+\alpha_x{\hat P}_x+\alpha_y{\hat P}_y+F_3\tau_3\beta\Sigma_3\ , 
\eeq
where, by using the Dirac representation of the Dirac gamma matrices, 
\beq\label{2-7}
& &
{\alpha}_i=\gamma^0\gamma^i=\left(
\begin{array}{cc}
0 & \sigma_i \\
\sigma_i & 0
\end{array}
\right)\ , \qquad
\beta=\gamma^0
=\left(
\begin{array}{cc}
1 & 0 \\
0 & -1 
\end{array}
\right)\ , \nonumber\\
& &
{\hat p}_i=-i\frac{\partial}{\partial x^i} \ , \qquad
{\hat P}_x={\hat p}_x+\frac{QB}{2}y\ , \qquad
{\hat P}_y={\hat p}_y-\frac{QB}{2}x\ . 
\eeq

Hereafter, let us consider two cases, namely $F>0$ and $F=0$, respectively.

\subsection{$F>0$ case}

The Dirac equation is written by $i\partial \psi/\partial t=h\psi=E\psi$, namely
\beq\label{2-8}
\left(
\begin{array}{cc}
F_3\tau_3\sigma_z-E & {\hat p}_z\sigma_z+{\hat P}_x\sigma_x+{\hat P}_y\sigma_y \\
{\hat p}_z\sigma_z+{\hat P}_x\sigma_x+{\hat P}_y\sigma_y & -F_3\tau_3\sigma_z-E 
\end{array}
\right)
\left(
\begin{array}{c}
\phi \\ 
\varphi
\end{array}
\right)=0\ , 
\eeq
where $\phi$ and $\varphi$ are two-component spinors. 
Eliminating $\varphi$ and noting $[\ {\hat P}_x\ , \ {\hat P}_y\ ]=iQB$, we obtain the following equation: 
\beq\label{2-9}
& &\biggl[\left\{
-E+\frac{E}{E^2-F^2}({\hat p}_z^2+{\hat P}_x^2+{\hat P}_y^2)-\frac{F}{E^2-F^2}QB\right\}\nonumber\\
& &\ +\sigma_z\left\{F-\frac{F}{E^2-F^2}({\hat p}_z^2-{\hat P}_x^2-{\hat P}_y^2)-\frac{E}{E^2-F^2}QB\right\}
\nonumber\\
& &\ \ 
-\frac{2F}{E^2-F^2}{\hat p}_z(\sigma_x{\hat P}_x+\sigma_y{\hat P}_y)\biggl]\phi=0\ , \\
& &\ \ \ F_3\tau_3=F\tau_f\tau_3
=\left(
\begin{array}{cc}
F & 0 \\
0 & F  
\end{array}\right) 
=F{\bf 1}
\ , \qquad
Q=\left\{
\begin{array}{ll}
\frac{2}{3}e & {\rm for\ up\ quark} \\
-\frac{1}{3}e & {\rm for\ down\ quark} 
\end{array}\right. 
\ ,
\nonumber
\eeq
where ${\bf 1}$ is the identity matrix in the isospin space. 
Let us consider $Q>0$ case, that is, the case of up quark. 
We introduce new operators instead of ${\hat P}_x$ and ${\hat P}_y$ as
\beq\label{2-10}
a=\frac{1}{\sqrt{2QB}}({\hat P}_x+i{\hat P}_y)\ , \qquad
a^{\dagger}=\frac{1}{\sqrt{2QB}}({\hat P}_x-i{\hat P}_y)\ ,
\eeq
where the commutation relation $[\ a\ , \ a^{\dagger}\ ]=1$ is satisfied. 
Then, Eq.(\ref{2-9}) is recast into 
\beq\label{2-11}
& &\biggl[\left\{
-E+\frac{E}{E^2-F^2}({\hat p}_z^2+QB(aa^{\dagger}+a^{\dagger}a)-\frac{F}{E^2-F^2}QB\right\}\nonumber\\
& &\ +\sigma_z\left\{F-\frac{F}{E^2-F^2}({\hat p}_z^2-QB(aa^{\dagger}+a^{\dagger}a)-\frac{E}{E^2-F^2}QB\right\}
\nonumber\\
& &\ \ 
-\frac{2F}{E^2-F^2}{\hat p}_z\sqrt{\frac{QB}{2}}(\sigma_x(a+a^{\dagger})-i\sigma_y(a-a^{\dagger}))\biggl]\phi=0\ .
\eeq
Here, expressing the two-component spinor $\phi$ as $\phi={}^t(\phi_1,\phi_2)$, 
the above equation can be expressed as 
\beq\label{2-11add}
& &\left[X_0+Y_0+(X_1+Y_1)a^{\dagger}a\right]\phi_1-Ca^{\dagger}\phi_2=0\ , \nonumber\\
& &-Ca\phi_1+\left[X_0-Y_0+(X_1-Y_1)a^{\dagger}a\right]\phi_2=0\ , 
\eeq
and eliminating $\phi_2$, 
the following equation can be derived from Eq.(\ref{2-11}):
\beq\label{2-12}
& &\left[
[X_0+Y_0+(X_1+Y_1)a^{\dagger}a]-[X_0-Y_0+(X_1-Y_1)(a^{\dagger}a-1)]^{-1}C^2a^{\dagger}a\right]\phi_1=0\ , 
\nonumber\\
& &
\eeq 
where we define the following quantity for simplicity:
\beq\label{2-13}
& &
X_0=-E+\frac{E{\hat p}_z^2}{E^2-F^2}+\frac{E-F}{E^2-F^2}QB\ , 
\qquad
Y_0=F-\frac{F{\hat p}_z^2}{E^2-F^2}-\frac{E-F}{E^2-F^2}QB\ , \nonumber\\
& &
X_1=\frac{2EQB}{E^2-F^2}\ , \qquad Y_1=\frac{2FQB}{E^2-F^2}\ , \qquad
C=\frac{2F{\hat p}_z\sqrt{2QB}}{E^2-F^2}\ . 
\eeq 
Here, ${\hat p}_z$ and $a^{\dagger}a$ should be replaced by their eigenvalues $p_z$ and $\nu\ (=0,1,2,\cdots)$. 
Further, since the case $Q>0$ is treated, namely the case of up quark, so $F_3=F$. 
Then from the coefficient of $\phi_1$ being 0, we can get the eigenvalue of the Dirac equation, $E$, which is expressed as 
$\epsilon_{p_z,\nu}^{\rm flavor}$, as 
\beq\label{2-14}
\epsilon_{p_z,\nu}^{\rm up}=\sqrt{(F\pm \sqrt{2QB\nu})^2+p_z^2} \qquad ({\rm for}\ \ \nu=1,\ 2, \cdots )
\eeq 
with $Q=2e/3$. 
It should be here noted that it is necessary to pay a special attention for the case $\nu=0$ because 
$a\phi({\nu=0})=a\ket{\nu=0}=0$ satisfies. 
For $\nu=0$, from $\phi_1=\ket{\nu=0}$, Eq.(\ref{2-11}) is reduced to 
\beq\label{2-11-2nd}
& &X_0+Y_0=0\ , \qquad \phi_2=0 
\eeq
because $a\phi_1=a\ket{\nu=0}=0$. 
Form $X_0+Y_0=0$, which leads to a quadratic equation with respect to $E$, we have 
\beq\label{2-15add}
\epsilon_{p_z,\nu=0}^{\rm up}=\sqrt{F^2+p_z^2}
\eeq
for positive energy solution. 
Thus, for $\nu=0$, the solution only appears once, and the energy corresponding to $\nu=0$ is not degenerate.

In the same way, we can get the eigenvalue $E$ in the case of down quark where $Q<0$. 
Instead of (\ref{2-10}), we define new operators as
\beq\label{2-15}
a=\frac{1}{\sqrt{-2QB}}({\hat P}_x-i{\hat P}_y)\ , \qquad
a^{\dagger}=\frac{1}{\sqrt{-2QB}}({\hat P}_x+i{\hat P}_y)\ , 
\eeq
where $[\ a\ , \ a^{\dagger}\ ]=1$ is satisfied. 
Then, we obtain the equation instead of (\ref{2-11}) as 
\beq\label{2-16}
& &\biggl[\left\{
-E+\frac{E}{E^2-F^2}({\hat p}_z^2-QB(aa^{\dagger}+a^{\dagger}a)-\frac{F}{E^2-F^2}QB\right\}\nonumber\\
& &\ +\sigma_z\left\{F-\frac{F}{E^2-F^2}({\hat p}_z^2+QB(aa^{\dagger}+a^{\dagger}a)-\frac{E}{E^2-F^2}QB\right\}
\nonumber\\
& &\ \ 
-\frac{2F}{E^2-F^2}{\hat p}_z\sqrt{-\frac{QB}{2}}(\sigma_x(a+a^{\dagger})+i\sigma_y(a-a^{\dagger}))\biggl]\phi=0\ .
\eeq
Eliminating $\phi_1$ which is the upper component of the two-component spinor $\phi$, we obtain the following equation:
\beq\label{2-17}
& &\left[
[{\tilde X}_0+{\tilde Y}_0-({\tilde X}_1+{\tilde Y}_1)a^{\dagger}a]-[{\tilde X}_0-{\tilde Y}_0
-({\tilde X}_1-{\tilde Y}_1)(a^{\dagger}a-1)]^{-1}{\tilde C}^2a^{\dagger}a\right]\phi_2=0\ , 
\nonumber\\
& &
\eeq 
where
\beq\label{2-18}
& &
{\tilde X}_0=-E+\frac{E{\hat p}_z^2}{E^2-F^2}-\frac{E+F}{E^2-F^2}QB\ , 
\qquad
{\tilde Y}_0=-F+\frac{F{\hat p}_z^2}{E^2-F^2}+\frac{E+F}{E^2-F^2}QB\ , \nonumber\\
& &
{\tilde X}_1=\frac{2EQB}{E^2-F^2}\ , \qquad {\tilde Y}_1=-\frac{2FQB}{E^2-F^2}\ , \qquad
{\tilde C}=-\frac{2F{\hat p}_z\sqrt{-2QB}}{E^2-F^2}\ . 
\eeq 
Thus, we obtain the eigenvalue for down quark, $\epsilon_{p_z,\nu}^{\rm down}$ by replacing ${\hat p}_z$ and $a^{\dagger}a$ into their eigenvalues 
$p_z$ and $\nu\ (=0,1,2,\cdots$):
\beq\label{2-19}
& &\epsilon_{p_z,\nu}^{\rm down}=\sqrt{(F\pm \sqrt{-2QB\nu})^2+p_z^2} \qquad ({\rm for}\ \ \nu=1,\ 2, \cdots)\nonumber\\
& &\epsilon_{p_z,\nu=0}^{\rm down}=\sqrt{F^2+p_z^2} \qquad\qquad\qquad\qquad ({\rm for}\ \ \nu=0)
\eeq
with $F_3=-F$ and $Q=-e/3$.

Thus, the single-particle energy of quarks with flavor $f=u$ or $d$ for up and down quarks can be expressed as 
\beq\label{2-20}
\epsilon_{p_z,\nu,\eta}^{f}=\sqrt{\left(F+\eta\sqrt{2|Q_f|B\nu}\right)^2+p_z^2}
\qquad
\left\{
\begin{array}{ll}
\nu=0,\ 1,\ 2, \cdots &{\rm for}\ \ \eta=1,\\
\nu=1,\ 2, \cdots & {\rm for}\ \ \eta=-1, \\
\end{array}\right. \quad\ \ 
\eeq
with $Q_u=2e/3$, $Q_d=-e/3$ and $\eta=\pm 1$.

The thermodynamic potential $\Phi$ can be expressed in terms of the vacuum expectation values of the Hamiltonian ${\hat H}$ and 
the particle number operator ${\hat N}$, in which the quarks occupy the energy levels from that with the lowest energy to that with the Fermi energy. 
Thus, we obtain $\Phi$ as
\beq\label{2-21}
\Phi&=&\frac{1}{V}\langle {\hat H}-\mu {\hat N}\rangle\nonumber\\
&=&3\cdot\frac{1}{V}\sum_{p_z,\nu\ (\epsilon_{p_z,\nu,+}^{f}\leq\mu)}[(\epsilon_{p_z,\nu,+}^{u}-\mu)+(\epsilon_{p_z,\nu,+}^{d}-\mu)]
\nonumber\\
& &
+3\cdot\frac{1}{V}\sum_{p_z,\nu\ (\epsilon_{p_z,\nu,-}^{f}\leq\mu)}[(\epsilon_{p_z,\nu,-}^{u}-\mu)+(\epsilon_{p_z,\nu,-}^{d}-\mu)]
+\frac{F^2}{2G}\ , 
\eeq
where $V$ represents the volume under consideration and the factor 3 represents the color degree of freedom.
Here, it should be noted that the sum with respect to $p_z$ and $\nu$ can be regarded as the following
\footnote{
Usually, $\frac{1}{V}\sum_{{\mib p}}$ is replaced to $\int\frac{d^3{\mib p}}{(2\pi)^3}$. 
Here, $p_x^2+p_y^2$ can be regarded as $2|Q|B\nu\ (=p_{\perp}^2)$. 
Thus, the following correspondence may be understood: 
$\int\!\!\int dp_x dp_y=\int 2\pi p_{\perp} dp_{\perp}=2\pi\int\sqrt{2|Q|B\nu}\frac{\sqrt{2|Q|B}}{2\sqrt{\nu}}d\nu=2\pi |Q|B\int d\nu=2\pi |Q|B\sum_{\nu}$.
}
:
\beq\label{2-22}
\frac{1}{V}\sum_{p_z,\nu}=\int\frac{dp_z}{2\pi}\cdot\frac{|Q_f|B}{2\pi}\sum_{\nu=\nu_{\rm min}^{f(\eta)}}^{\nu_{\rm max}^{f(\eta)}}\ , 
\eeq 
Thus, the thermodynamic potential can be obtained finally as 
\beq\label{2-23}
& &\Phi=3\int_{-p_F}^{p_F}\frac{dp_z}{2\pi}\sum_{f=u,d,\eta=\pm}\frac{|Q_f|B}{2\pi}\sum_{\nu=\nu_{\rm min}^{f(\eta)}}^{\nu_{\rm max}^{f(\eta)}}
\left[\sqrt{\left(F+\eta\sqrt{2|Q_f|B\nu}\right)^2+p_z^2}-\mu\right]+\frac{F^2}{2G}\ , 
\nonumber\\
& &
\eeq 
where $p_F$, $\nu_{\rm min}^{f(\eta)}$ and $\nu_{\rm max}^{f(\eta)}$ are determined by the condition $\epsilon_{p_z,\nu,\eta}^{f}\leq\mu$.

\subsection{$F=0$ case}

Next, let us consider the $F=0$ case. 
For $Q>0$ and $Q<0$, Eqs.(\ref{2-11}) and (\ref{2-16}) are valid with $F=0$. 
Thus, the following equation for the two-component spinor $\phi$ is obtained: 
\beq\label{2-24}
\left[\left\{-E+\frac{1}{E}({\hat p}_z^2+|Q|B(2a^{\dagger}a+1)\right\}\mp \sigma_z\frac{1}{E}|Q|B\right]\phi=0\ , 
\eeq
where the upper (lower) sign in front of $\sigma_z$ corresponds to the case $Q>0$ ($Q<0$). 
Replacing ${\hat p}_z$ and $a^{\dagger}a$ by their eigenvalues and $\phi={}^t(\phi_1,\phi_2)$, 
the above equation is written as 
\beq\label{2-25}
& &\left(-E^2+p_z^2+2|Q|B\nu\right)\phi_{\!\!\!{\tiny {\begin{array}{c} 1 \\ 2 \end{array}}}}=0\ , \nonumber\\
& &\left(-E^2+p_z^2+2|Q|B(\nu+1)\right)\phi_{\!\!\!{\tiny {\begin{array}{c} 2 \\ 1 \end{array}}}}=0\ .
\eeq 
Thus, the single-particle energy $E$ is obtained as 
\beq\label{2-26}
E=
\left\{
\begin{array}{l}
\sqrt{p_z^2+2|Q|B\nu} \\
\sqrt{p_z^2+2|Q|B(\nu+1)}
\end{array}
\right. \ . 
\eeq
This result is included in Eq.(\ref{2-20}).
Thus, the thermodynamic potential $\Phi_0$ can be calculated as 
\beq\label{2-27}
\Phi_0&=&
3\sum_{f=u,d}\int_{-p_0}^{p_0}\frac{dp_z}{2\pi}\cdot\frac{|Q_f|B}{2\pi}\Bigg[
\sum_{\nu=0}^{\nu_{\rm M}^{f}}
\left(\sqrt{p_z^2+2|Q_f|B\nu}-\mu\right)
\nonumber\\
& &\qquad\qquad\qquad\qquad\qquad\qquad
+\sum_{\nu=0}^{\nu_{\rm M}^{f}-1}
\left(\sqrt{p_z^2+2|Q_f|B(\nu+1)}-\mu\right)\Bigg]\nonumber\\
&=&3\sum_{f=u,d}\int_{-p_0}^{p_0}\frac{dp_z}{(2\pi)^2}|Q_f|B(|p_z|-\mu)
\nonumber\\
& & 
+3\sum_{f=u,d}\int_{-p_0}^{p_0}\frac{dp_z}{(2\pi)^2}2|Q_f|B\sum_{\nu=1}^{\nu_{\rm M}^f}\left[\sqrt{p_z^2+2|Q_f|B\nu}-\mu\right]\ ,
\eeq
where $p_0$ and $\nu_{\rm M}^f$ should be determined later.

\setcounter{equation}{0}
\section{The thermodynamic potential in three cases: $F>\mu$, $0<F<\mu$ and $F=0$}

First, let us integrate out with respect to $p_z$ in the thermodynamic potential (\ref{2-23}) and (\ref{2-27}). 
The condition $\left(F+\eta\sqrt{2|Q_f|B\nu}\right)^2+p_z^2 \leq \mu^2$ gives the range of integration with respect to $p_z$. 
Namely, 
\beq\label{3-1}
-\sqrt{\mu^2-\left(F+\eta\sqrt{2|Q_f|B\nu}\right)^2}\leq p_z \leq
\sqrt{\mu^2-\left(F+\eta\sqrt{2|Q_f|B\nu}\right)^2}\ (\equiv p_F)\ . \quad
\eeq
By using the following integration formula,  
\beq\label{3-2}
& &\int\!\! dp_z\sqrt{(F+X)^2+p_z^2}=\frac{p_z}{2}\sqrt{(F+X)^2+p_z^2}+\frac{(F+X)^2}{2}\ln \left(p_z+\sqrt{(F+X)^2+p_z^2}\right) , 
\nonumber\\
& &
\eeq
we can carry out the integration with respect to $p_z$ in Eq.(\ref{2-23}), which leads to 
\beq\label{3-3}
& &
\Phi=\frac{3}{2\pi}\sum_{f=u,d; \eta=\pm}
\frac{|Q_f|B}{2\pi}\sum_{\nu=\nu_{f,m}^{(\eta)}}^{\nu_{f,M}^{(\eta)}}
\biggl[
-\mu\sqrt{\mu^2-\left(F+\eta\sqrt{2|Q_f|B\nu}\right)^2}
\nonumber\\
& &\qquad\qquad\qquad\qquad\qquad\qquad
+\left(F+\eta\sqrt{2|Q_f|B\nu}\right)^2\ln\frac{\mu+\sqrt{\mu^2-\left(F+\eta\sqrt{2|Q_f|B\nu}\right)^2}}{F+\eta\sqrt{2|Q_f|B\nu}}
\biggl]
\nonumber\\
& &\qquad\qquad\qquad
+\frac{F^2}{2G}\nonumber\\
& &\qquad
=\frac{F^2}{2G}+\frac{3}{8\pi^2}\sum_{f=u,d;\eta=\pm}
2|Q_f|B\sum_{\nu=\nu_{f,m}^{(\eta)}}^{\nu_{f,M}^{(\eta)}}g_{\eta}(2|Q_f|B\nu)
\ ,
\eeq
where $\nu_{f,m}^{(\eta)}$ and $\nu_{f,M}^{(\eta)}$ are determined by the condition 
\beq\label{3-4}
\mu^2-\left(F+\eta\sqrt{2|Q_f|B\nu}\right)^2 \geq 0\ , 
\eeq 
which guarantees that $p_z$ is real. 
Here, $g_{\eta}(x)$ is defined by 
\beq\label{3-5}
g_{\eta}(x)=-\mu\sqrt{\mu^2-(F+\eta\sqrt{x})^2}
+(F+\eta\sqrt{x})^2\ln
\frac{\mu+\sqrt{\mu^2-(F+\eta\sqrt{x})^2}}{F+\eta\sqrt{x}}\ .
\eeq

\subsection{$F\geq \mu$ case}

For $\eta=1$, the condition (\ref{3-4}) is not satisfied for any $F$. 
On the other hand, for $\eta=-1$, if $F<\sqrt{2|Q_f|B\nu}$, the condition (\ref{3-4}) gives $\sqrt{2|Q_f|B\nu}\leq\mu+F$. 
If $F>\sqrt{2|Q_f|B\nu}$, the condition (\ref{3-4}) gives $\sqrt{2|Q_f|B\nu}\geq F-\mu$. 
Thus, we summarize the condition for $\nu$ as follows: 
\beq\label{3-6}
0\neq \nu_{f,m}^{(-)}\equiv 
\left[\frac{(F-\mu)^2}{2|Q_f|B}\right]+1 \leq \nu 
\leq \left[\frac{(F+\mu)^2}{2|Q_f|B}\right]\equiv \nu_{f,M}^{(-)}\ ,
\eeq
where $[\cdots ]$ represents the Gauss symbol. 
Thus, the thermodynamic potential can be expressed as 
\beq\label{3-7}
\Phi&=&\frac{3}{2\pi}\sum_{f=u,d}
\frac{|Q_f|B}{2\pi}\sum_{\nu=\nu_{f,m}^{(-)}}^{\nu_{f,M}^{(-)}}
\biggl[
-\mu\sqrt{\mu^2-\left(F-\sqrt{2|Q_f|B\nu}\right)^2}
\nonumber\\
& &\qquad\qquad\qquad\qquad\qquad\qquad
+\left(F-\sqrt{2|Q_f|B\nu}\right)^2\ln\frac{\mu+\sqrt{\mu^2-\left(F-\sqrt{2|Q_f|B\nu}\right)^2}}{F-\sqrt{2|Q_f|B\nu}}
\biggl]
\nonumber\\
& &\qquad\qquad\qquad
+\frac{F^2}{2G}
\nonumber\\
&=&\frac{3}{8\pi^2}\sum_{f=u,d}2|Q_f|B\sum_{\nu=\nu_{f,m}^{(-)}}^{\nu_{f,M}^{(-)}}
g_-(2|Q_f|B\nu)+\frac{F^2}{2G}
\eeq
with (\ref{3-6}).

\subsection{$F < \mu$ case}

For $\eta=1$, the condition (\ref{3-4}) gives the condition 
$0 \leq \sqrt{2|Q_f|B\nu}\leq \mu-F$. 
On the other hand, for $\eta=-1$, if $F<\sqrt{2|Q_f|B\nu}$, the condition (\ref{3-4}) gives $F\leq \sqrt{2|Q_f|B\nu}\leq\mu+F$. 
If $F>\sqrt{2|Q_f|B\nu}$, the condition (\ref{3-4}) gives $0\leq\sqrt{2|Q_f|B\nu}\leq F$. 
Thus, we summarize the condition for $\nu$ as follows: 
\beq\label{3-8}
& &\nu_{f,m}^{(+)}\equiv 
0 \leq \nu 
\leq \left[\frac{(\mu-F)^2}{2|Q_f|B}\right]\equiv \nu_{f,M}^{(+)} \qquad {\rm for}\quad \eta=1\ , 
\nonumber\\
& &
\nu_{f,m}^{(-)}\equiv 
1 \leq \nu 
\leq \left[\frac{(\mu+F)^2}{2|Q_f|B}\right]\equiv \nu_{f,M}^{(-)} \qquad {\rm for}\quad \eta=-1\ . 
\eeq
Thus, the thermodynamic potential is (\ref{3-3}) with (\ref{3-8}) for $\eta=\pm$.  
Namely, 
\beq\label{3-9}
& &
\Phi
=\frac{F^2}{2G}+\frac{3}{8\pi^2}\sum_{f=u,d;\eta=\pm}
2|Q_f|B\sum_{\nu=\nu_{f,m}^{(\eta)}}^{\nu_{f,M}^{(\eta)}}g_{\eta}(2|Q_f|B\nu)
\ .
\eeq

\subsection{$F=0$ case}

Next, let us consider $F=0$ case. 
Thus, the thermodynamic potential $\Phi_0$ can be given in (\ref{2-27}), which we 
show again: 
\beq\label{3-10}
\Phi_0&=&
3\sum_{f=u,d}\int_{-\mu}^{\mu}\frac{dp_z}{(2\pi)^2}|Q_f|B(|p_z|-\mu)\nonumber\\
& &+3\sum_{f=u,d}\int_{-\sqrt{\mu^2-2|Q_f|B\nu}}^{\sqrt{\mu^2-2|Q_f|B\nu}}
\frac{dp_z}{(2\pi)^2}
2|Q_f|B\sum_{\nu=1}^{\nu_{\rm M}^f}\left[\sqrt{p_z^2+2|Q_f|B\nu}-\mu\right]\nonumber\\
&=&
-\frac{3}{4\pi^2}\sum_{f=u,d}|Q_f|B\mu^2
\nonumber\\
& &
+\frac{3}{4\pi^2}\sum_{f=u,d}2|Q_f|B
\sum_{\nu=1}^{\nu_{\rm M}^f}\left[
-\mu\sqrt{\mu^2-2|Q_f|B\nu}+2|Q_f|B\nu
\ln\frac{\mu+\sqrt{\mu^2-2|Q_f|B\nu}}{\sqrt{2|Q_f|B\nu}}\right]\nonumber\\
&=&
-\frac{3}{4\pi^2}\sum_{f=u,d}|Q_f|B\mu^2
+\frac{3}{4\pi^2}\sum_{f=u,d}2|Q_f|B
\sum_{\nu=1}^{\nu_{\rm M}^f}
g_0(2|Q_f|B\nu)\ , \nonumber\\
& &g_0(x)=
-\mu\sqrt{\mu^2-x}
+x\ln\frac{\mu+\sqrt{\mu^2-x}}{\sqrt{x}}
\eeq
with
\beq\label{3-11}
\nu_{\rm M}^f=\left[\frac{\mu^2}{2|Q_f|B}\right]\ .
\eeq

\setcounter{equation}{0}
\section{Approximation of the thermodynamic potential by replacing 
summation by integration with respect to the Landau level}

In the thermodynamic potential (\ref{2-23}) and (\ref{2-27}), 
the quantum number $\nu$, which labels the Landau level, has to be summed up.  
However, since it is interesting to consider the spontaneous magnetization, it may be assumed that the external magnetic field $B$ 
is small and finally $B$ becomes 0.
Therefore, let us replace the sum with respect to $\nu$ by an integration approximately \cite{BJ}. 

In general, let us consider a function $f(x)$. 
Here, we introduce a small quantity $a$ and let us consider the Tailor expansion around $x=a\nu$ as follows:
\beq\label{4-1}
\int_{a(\nu-1)}^{a\nu}dx f(x)
&=&
\int_{a(\nu-1)}^{a\nu}dx \left[
f(a\nu)+\left.\frac{df}{dx}\right|_{x=a\nu}(x-a\nu)+\left.\frac{1}{2}\frac{d^2 f}{dx^2}\right|_{x=a\nu}(x-a\nu)^2+\cdots\right]
\nonumber\\
&=&af(a\nu)-\frac{1}{2}a^2f'(a\nu)+\frac{1}{6}a^3f''(a\nu)+\cdots\ .
\eeq 
Thus, the following relations is obtained :
\beq\label{4-2}
\sum_{\nu=\nu_m+1}^{\nu_M}\int_{a(\nu-1)}^{a\nu}dx f(x)\!\!\! &\equiv&\!\!\! \int_{a\nu_m}^{a\nu_M}dx f(x)\nonumber\\
&=&\!\!\!
a\!\!\sum_{\nu=\nu_m+1}^{\nu_M}\!\!f(a\nu)-\frac{a^2}{2}\!\!\sum_{\nu=\nu_m+1}^{\nu_M}\!\!f'(a\nu)+\frac{a^3}{6}\!\!\sum_{\nu=\nu_m+1}^{\nu_M}\!\!f''(a\nu)+\cdots\ .\ \qquad 
\eeq
Here, it should be noted that the definition of integral can be used when $a$ is infinitesimally small, namely, 
\beq\label{4-3}
& &a\sum_{\nu=\nu_m+1}^{\nu_M}f'(a\nu)=\int_{a\nu_m}^{a\nu_M}dx f'(x) =f(a\nu_M)-f(a\nu_m)\ , 
\eeq
and so on. 
Thus, adding $af(a\nu_m)$ on both sides of Eq.(\ref{4-2}), a useful approximate formula is obtained as follows:
\beq\label{4-4}
& &a\sum_{\nu=\nu_m}^{\nu_M}f(a\nu)
=\int_{a\nu_m}^{a\nu_M}dx f(x)+\frac{a}{2}\left[f(a\nu_M)+f(a\nu_m)\right]
-\frac{a^2}{6}\left[f'(a\nu_M)-f'(a\nu_m)\right]+\cdots\ .
\nonumber\\
& &
\eeq
After here, let us approximate the thermodynamic potential in the case of small $B$.

For $F>\mu$, Eq.(\ref{3-7}) is approximated by using the formula (\ref{4-4}) as follows:
\beq\label{4-5}
\Phi&=&\frac{F^2}{2G}
+
\frac{3}{8\pi^2}
\sum_{f=u,d}
\biggl[
2\int_{p_{\rm min}^f}^{p_{\rm max}^f}dp_{\perp}p_{\perp}g_-(p_{\perp}^2)\nonumber\\
& &\qquad\qquad\qquad\qquad
+|Q_f|B\left(g_-(2|Q_f|B\nu_{f,m}^{(-)})+g_-(2|Q_f|B\nu_{f,M}^{(-)})\right)\nonumber\\
& &\qquad\qquad\qquad
-\frac{2|Q_F|^2B^2}{3}\left(
g_-'(2|Q_f|B\nu_{f,M}^{(-)})-g_-'(2|Q_f|B\nu_{f,m}^{(-)})\right)+\cdots\biggl]\quad
\eeq
with (\ref{3-5}). 
Here, $p_{\rm min}^f=\sqrt{2|Q_f|B\nu_{f,m}^{(-)}}$
and
$p_{\rm max}^f=\sqrt{2|Q_f|B\nu_{f,M}^{(-)}}$, respectively.

For $F<\mu$, the thermodynamic potential (\ref{3-9}) is approximated by using (\ref{4-2}) directly as
\beq\label{4-6}
\Phi&=&
\frac{F^2}{2G}+\frac{3}{8\pi^2}\sum_{f=u,d}2|Q_f|B\left[
g_+(0)+\sum_{\nu=1}^{\nu_{f,M}^{(+)}}g_+(2|Q_f|B\nu)+\sum_{\nu=1}^{\nu_{f,M}^{(-)}}g_-(2|Q_f|B\nu)\right]
\nonumber\\
&=&\frac{F^2}{2G}+\frac{3}{8\pi^2}\sum_{f=u,d}2|Q_f|Bg_+(0)
\nonumber\\
& &+
\frac{3}{8\pi^2}
\sum_{f=u,d;\eta=\pm}
\biggl[
2\int_{0}^{p_{\rm max}^{f(\eta)}}dp_{\perp}p_{\perp}g_{\pm}(p_{\perp}^2)
+|Q_f|B\left(g_{\pm}(2|Q_f|B\nu_{f,M}^{(\eta)})-g_{\eta}(0)\right)\nonumber\\
& &\qquad\qquad\qquad\qquad
-\frac{2|Q_f|^2B^2}{3}\left(g_{\eta}'(2|Q_f|B\nu_{f,M}^{(\eta)})-g_{\eta}'(0)\right)+\cdots\biggl]\ , 
\eeq
where $\nu_{f,M}^{(\eta)}=[(\mu\mp F)^2/(2|Q_f|B)]$ and 
$p_{\rm max}^{f(\eta)}=\sqrt{2|Q_f|B\nu_{f,M}^{(\eta)}}$, respectively.

For $F=0$, the thermodynamic potential (\ref{3-10}) is approximated as 
\beq\label{4-7}
\Phi_0&=&-\frac{3}{4\pi^2}\sum_{f=u,d}|Q_f|B\mu^2
\nonumber\\
& &+\frac{3}{4\pi^2}\sum_{f=u,d}\biggl[
\int_0^{a_f\nu_M^f}g_0(x)dx+|Q_f|B\left(g_0(a_f\nu_M^f)-g_0(0)\right)\nonumber\\
& &\qquad\qquad\qquad
-\frac{2|Q_f|^2B^2}{3}\left(g_0'(2|Q_f|B\nu_M^f)-g_0'(0)\right)+\cdots\biggl]\ . 
\eeq

Here, since $B$ is small, $\nu$ may be regarded as a continuum variables. 
Remembering the condition which determines $\nu_{f,m}^{(\eta)}$ and $\nu_{f,M}^{(\eta)}$, 
namely, $g_{\pm}(2|Q_f|B\nu_{f,M}^{(\eta)})=0$, and $g_{\pm}(0)\neq 0$ with (\ref{3-5}), then Eqs.(\ref{4-5}),  (\ref{4-6}) and (\ref{4-7}) are simply calculated by 
performing the integration as
\beq\label{4-8}
& &\Phi=\Phi_>=-\frac{F\mu^3}{2\pi}+\frac{F^2}{2G}+O(B^2)\ , \qquad {\rm for}\quad F>\mu\ , \nonumber\\
& &\Phi=\Phi_<=\frac{3}{\pi^2}\biggl[\sqrt{\mu^2-F^2}\left(\frac{\mu^3}{6}+\frac{F^2\mu}{4}\right)-\frac{F\mu^3}{3}\arctan\frac{F}{\sqrt{\mu^2-F^2}}
\nonumber\\
& &\qquad\qquad\qquad
+\frac{F^4}{12}\ln \frac{\mu+\sqrt{\mu^2-F^2}}{F}\biggl]+\frac{F^2}{2G}
+O(B^2)\ , 
\qquad {\rm for}\quad F<\mu\ , \nonumber\\
& &\Phi=\Phi_0=-\frac{\mu^4}{2\pi^2}+O(B^2)\ , \qquad {\rm for}\quad F=0\ ,
\eeq
where we used the integration formulae:
\beq\label{4-9}
& &\int dy y\sqrt{\mu^2-(F+\eta y)^2}
=\sqrt{\mu^2-(F+\eta y)^2}\left[
-\frac{\mu^2}{3}-\frac{F}{2}\left(F+\eta y\right)+\frac{1}{3}\left(F+\eta y\right)^2\right]\nonumber\\
& &\qquad\qquad\qquad\qquad\qquad\qquad
-\frac{F\mu^2}{2}\arctan\left[
\frac{(F+\eta y)^2\sqrt{\mu^2-(F+\eta y)^2}}{\mu^2-(F+\eta y)^2}\right]\ , \nonumber\\
& &\int dy y (F+\eta y)^2\ln\frac{\mu+\sqrt{\mu^2-(F+\eta y)^2}}{F+\eta y}\nonumber\\
& &\qquad\qquad\qquad\qquad\qquad
=\sqrt{\mu^2-(F+\eta y)^2}\left[-\frac{\mu^3}{6}+\frac{F\mu}{6}(F+\eta y)-\frac{\mu}{12}(F+\eta y)^2\right]\nonumber\\
& &\qquad\qquad\qquad\qquad\qquad\ \ 
+\frac{F\mu^3}{6}\arctan\left[\frac{(F+\eta y)\sqrt{\mu^2-(F+\eta y)^2}}{-\mu^2+(F+\eta y)^2}\right]
\nonumber\\
& &\qquad\qquad\qquad\qquad\qquad\ \ 
+\frac{1}{12}(F+\eta y)^3\left[-4F+3(F+\eta y)\right]\ln\frac{\mu+\sqrt{\mu^2-(F+\eta y)^2}}{F+\eta y}\ . \nonumber\\
& &
\eeq

\setcounter{equation}{0}
\section{Spontaneous magnetization for the quark matter under the tensor-type interaction 
between quarks}

%
\begin{figure}[b]
\begin{center}
\includegraphics[height=5.5cm]{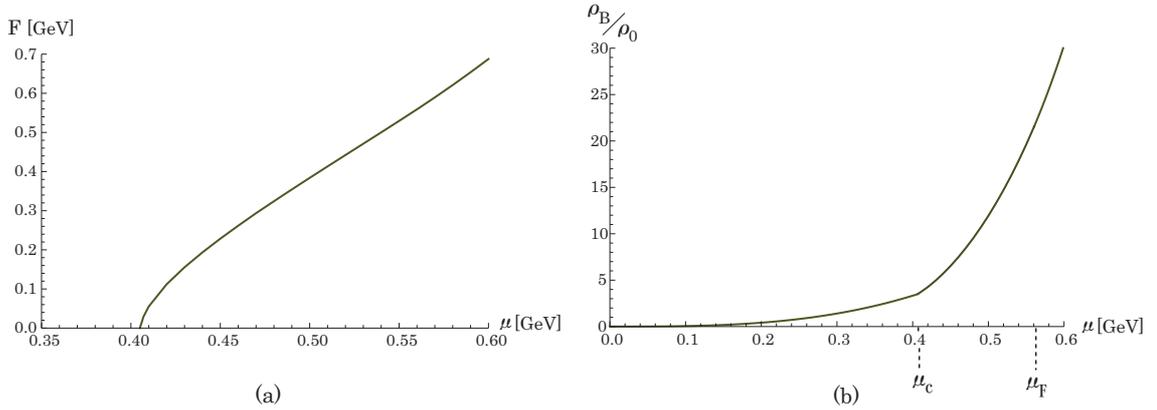}
\caption{(a) The spin polarization condensate $F$ is depicted as a function of the quark chemical potential $\mu$. (b) The baryon number density of quark matter divided by the normal nuclear density 
$\rho_0=0.17$ fm$^{-3}$ is shown as a function of the quark chemical potential. 
Here, $\mu_c\approx 0.407$ GeV is the value at which the spin polarization occurs. 
Also, in the region of $\mu\geq \mu_F\approx 0.5605$ GeV, the spin polarization condensate 
is greater than the chemical potential, $F\geq \mu$.
}
\label{fig:fig1}
\end{center}
\end{figure}
%
First, let us calculate the spin polarization $F$ with $B=0$ as a function of the chemical potential $\mu$. 
This task has been done in our previous papers \citen{oursPTP} or \citen{oursPTEP1}. 
The thermodynamic potential can be expressed as 
\beq\label{5-1}
& &\Phi_{B=0}=6\cdot\frac{1}{V}\sum_{{\mib p}(\epsilon_{{\mib p}^{(+)}\leq \mu})}\left(\epsilon_{\mib p}^{(+)}-\mu\right)
+6\cdot\frac{1}{V}\sum_{{\mib p}(\epsilon_{{\mib p}^{(-)}}\leq\mu)}\left(\epsilon_{\mib p}^{(-)}-\mu\right)
+\frac{F^2}{2G}\ , \\
& &\epsilon_{{\mib p}}^{(\pm)}=\sqrt{p_3^2+\left(F+\eta\sqrt{p_1^2+p_2^2}\right)^2}\ , \nonumber
\eeq
where the factor 6 represents the color and flavor degrees of freedom and 
$p_1=p_x$, $p_2=p_y$ and $p_3=p_z$. 
Here, $(1/V)\cdot\sum_{\mib p}$ can be replaced to the integration $\int d^3{\mib p}/(2\pi)^3$. 
Integrating the three-momentum, 
we have obtained the thermodynamic potential as
\beq\label{5-2}
& &\Phi_{B=0}=\left\{
\begin{array}{ll}
\displaystyle \frac{F^2}{2G}-\frac{1}{\pi^2}\biggl[
\frac{\sqrt{\mu^2-F^2}}{4}(3F^2\mu+2\mu^3)+F\mu^3\arctan\frac{F}{\sqrt{\mu^2-F^2}} \\
\displaystyle \qquad\qquad \ \ 
-\frac{F^4}{4}\ln\frac{\mu+\sqrt{\mu^2-F^2}}{F} & {\rm for}\quad F<\mu\\
\displaystyle \frac{F^2}{2G}-\frac{F\mu^3}{2\pi} & {\rm for}\quad F>\mu\ .
\end{array}\right. 
\nonumber\\
& &
\eeq
It should be here noted that Eq.(\ref{5-2}) is identical with (\ref{4-8}) with $B=0$. 
The gap equation is derived by $\partial \Phi_{B=0}/\partial F=0$. 
The quark number density $\rho_q$ is also derived through the thermodynamic relation 
$\rho_q=-\partial \Phi/\partial \mu$ \cite{oursPTP} for the solution of the gap equation, 
$F=F_{\rm min}$:
\beq\label{5-3}
& &
\rho_q=
\left\{
\begin{array}{ll}
\displaystyle \frac{1}{\pi^2}\biggl[
(F_{\rm min}^2+2\mu^2)\sqrt{\mu^2-F_{\rm min}^2}+3F_{\rm min}\mu^2\arctan\frac{F_{\rm min}}{\sqrt{\mu^2-F_{\rm min}^2}} \biggl] & {\rm for} \quad F<\mu\\
\displaystyle \frac{3G}{4\pi^2}\mu^5 & {\rm for} \quad F>\mu\ .
\end{array}\right.
\nonumber\\
& &
\eeq
Figure {\ref{fig:fig1}} shows (a) the spin polarization condensate $F$ and (b) 
the baryon density of quark matter divided by the normal nuclear density, 
$\rho_0=0.17$ fm${}^{-3}$, as a function of the quark chemical potential $\mu$. 
Here, we used a parameter $G=20$ GeV${}^{-2}$ which has been used
\footnote{
If the vacuum polarization is taken into account, the parameter $G$ with 
rather small value gives the same results quantitatively such as $G=11.1$ GeV${}^{-2}$ 
under the standard three-momentum cutoff $\Lambda=0.631$ GeV, 
although we do not consider the vacuum polarization in this paper.
}
in our previous papers \cite{oursPTP,oursPTEP1}.
About $\mu=\mu_c\approx 0.407$ GeV which corresponds to the baryon density being 3.53 $\rho_0$, 
the spin polarization appears against the free quark phase. 
Of course, if the two-flavor color superconductivity is considered, it has been shown that the spin polarized phase appears 
about $\mu=0.442$ GeV ($\rho_B\approx 5.85 \rho_0$)\cite{oursPTEP1}.

We derive the spontaneous magnetization through the thermodynamic relation. 
The spontaneous magnetization per unit volume, ${\cal M}$, is defined by 
\beq\label{5-4}
{\cal M}=-\left.\frac{\partial \Phi}{\partial B}\right|_{B=0}\ .
\eeq
In the right-hand side, $B=0$ is adopted. 
However, from (\ref{4-8}), the thermodynamic potential does not depend on $B$ linearly. 
Thus, the spontaneous magnetization is equal to zero under this consideration, namely
\beq\label{5-5}
{\cal M}=-\left.\frac{\partial \Phi}{\partial B}\right|_{B=0}=0\ , 
\eeq
even if the quark-spin polarization occurs, while the magnetic polarization which is proportional to $B$ may appear.

\vspace{0.8cm}
\setcounter{equation}{0}
\section{Spontaneous magnetization originated from the anomalous magnetic moments of quarks}

In the previous section it has been found that no spontaneous magnetization 
of polarized high density quark matter is predicted by the normal coupling to an external magnetic field. 
However, the spin polarization $F$ should be observed in some way. 
We know that the quark has an anomalous magnetic moment. 
In this section, let us investigate the effect of the anomalous magnetic moment of quarks.

The effect of the anomalous magnetic moment $\mu_A$ is introduced 
at the level of the mean field approximation in this paper. 
Here, it is known that the anomalous magnetic moments reveal as \cite{AMM} 
\beq\label{6-1}
& &\mu_A{\bf 1}=
\left(
\begin{array}{cc}
\mu_u & 0 \\
0 & \mu_d
\end{array}
\right)\ , \qquad
\mu_u=1.85\ \mu_N\ , \quad
\mu_d=-0.97\ \mu_N\ , \nonumber\\
& &\qquad\quad\qquad\qquad\qquad\qquad
\mu_N=\frac{e}{2m_p}=3.15\times 10^{-17} \ {\rm GeV/T}\ . 
\eeq
We introduce the effects of the anomalous magnetic moment in the Lagrangian density within the mean field approximation 
as 
\beq\label{6-2}
{\cal L}_A={\cal L}-\frac{i}{2}{\bar \psi}\mu_A\gamma^{\mu}\gamma^{\nu}F_{\mu\nu}\psi\ , 
\eeq
where $F_{\mu\nu}=\partial_{\mu}A_{\nu}-\partial_{\nu}A_{\mu}$ and
$F_{12}\equiv -B_z=-B$. 
Here, ${\cal L}$ is nothing but Eq.(\ref{2-3}). 
We only take $\mu$ and $\nu$ as $\mu=1,\ \nu=2$ and $\mu=2,\ \nu=1$ because 
the magnetic field has only $z$-component.  
Then, in the mean field approximation, Eq.(\ref{2-3}) is recast into 
\beq\label{6-3}
{\cal L}_A=i{\bar \psi}\gamma^\mu D_{\mu}\psi 
-{\bar \psi}(F_3\tau_3+\mu_A B{\bf 1})\Sigma_3\psi\ .
\eeq
Here, $F_3\tau_3=F{\bf 1}$ where ${\bf 1}$ is the $2 \times 2$ identity matrix for the isospin space, 
which are denoted in Eqs.(\ref{2-2}) and (\ref{2-9}).
Thus, we introduce the flavor-dependent variables ${\wtilde F}_f$ as 
\beq\label{6-4}
{\wtilde F}_f=F+\mu_f B\ , \quad{\rm namely}\quad 
{\wtilde F}_u=F+\mu_u B\ , \qquad
{\wtilde F}_d=F+\mu_d B\ .
\eeq
Therefore, the Lagrangian density can be expressed as 
\beq\label{6-5}
{\cal L}_A&=&i{\bar \psi}\gamma^{\mu}D_{\mu}\psi-{\bar \psi}(F+\mu_A B)\Sigma_3\psi
-\frac{F^2}{2G}\nonumber\\
&=&i{\bar \psi}\gamma^{\mu}D_{\mu}\psi-{\bar \psi}{\wtilde F}\Sigma_3\psi
-\frac{F^2}{2G}
\ . 
\eeq
Thus, we learn that the variable $F$ should be replaced to ${\wtilde F}_f=F+\mu_f B$ 
for each flavor except for the last term in (\ref{6-5}).
In this replacement, we can derive the thermodynamic potential in the same way developed 
in \S 4. 
Namely, the results including the effects of the anomalous magnetic moment of quarks 
should be obtained by replacing $F$ into ${\wtilde F}_f$ for each 
flavor in the previous calculation developed in \S 4, except for the 
term originated from the mean field approximation, $F^2/2G$.   
The results are summarized as follows: 
\beq\label{6-6}
& &\Phi=\Phi_>=-\frac{1}{2}\sum_{f=u,d}\frac{{\wtilde F}_f\mu^3}{2\pi}+\frac{F^2}{2G}+O(B^2)\ , \qquad {\rm for}\quad F>\mu\ , \nonumber\\
& &\Phi=\Phi_<=\frac{1}{2}\sum_{f=u,d}\frac{3}{\pi^2}\biggl[-\sqrt{\mu^2-{\wtilde F}_f^2}
\left(\frac{\mu^3}{6}+\frac{{\wtilde F}_f^2\mu}{4}\right)-
\frac{{\wtilde F}_f\mu^3}{3}\arctan\frac{{\wtilde F}_f}{\sqrt{\mu^2-{\wtilde F}_f^2}}
\nonumber\\
& &\qquad\qquad\qquad
+\frac{{\wtilde F}_f^4}{12}\ln \frac{\mu+\sqrt{\mu^2-{\wtilde F}_f^2}}{{\wtilde F}_f}\biggl]
+\frac{F^2}{2G}\ , \qquad {\rm for}\quad \mu\geq F>0\ ,
\eeq

\begin{figure}[t]
\begin{center}
\includegraphics[height=6cm]{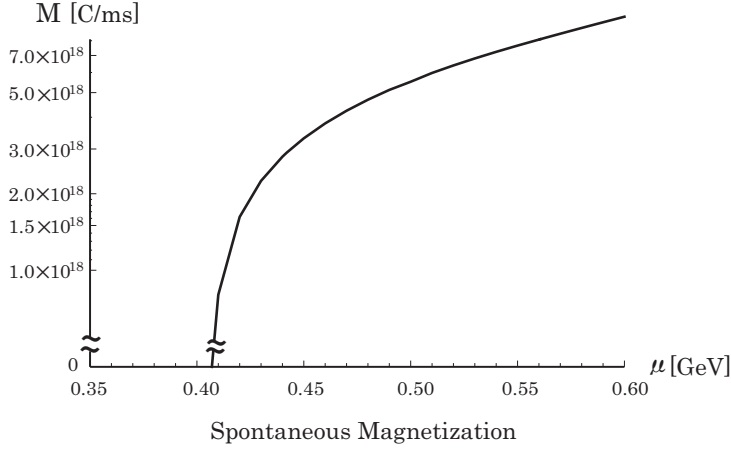}
\caption{
The spontaneous magnetization per unit volume ${\cal M}$ is depicted as a function of the quark chemical potential $\mu$. 
The vertical axis represents the magnetization ${\cal M}$ in a 
logarithmic scale.
For $\mu<0.407$ GeV, the magnetization does not occur because the spin polarization does not appear, $F=0$.
}
\label{fig:fig2}
\end{center}
\end{figure}

Thus, the spontaneous magnetization per unit volume, ${\cal M}$, can be 
derived thorough Eq.(\ref{5-4}), that is
\beq\label{6-7}
{\cal M}=-\left.\frac{\partial \Phi}{\partial B}\right|_{B=0}
=-\sum_{f=u,d}\left.\frac{\partial \Phi}{\partial {\wtilde F}_f}\right|_{B=0}
\cdot\frac{\partial {\wtilde F}_f}{\partial B}\ . 
\eeq
From (\ref{6-6}), we can derive the spontaneous magnetization originated form the 
anomalous magnetic moment and the spin polarization: 
\bsub\label{6-8}
\beq
{\cal M}&=&-\left.\frac{\partial \Phi_>}{\partial B}\right|_{B=0}
=\frac{\mu^3}{4\pi}(\mu_u+\mu_d)\ , 
\label{6-8a}\\
{\cal M}&=&-\sum_{f=u,d}\left.\frac{\partial \Phi_<}{\partial {\wtilde F}_f}\right|_{B=0}
\cdot\frac{\partial {\wtilde F}_f}{\partial B}
\nonumber\\
&=&
\frac{1}{2\pi^2}\left[
2F\mu\sqrt{\mu^2-F^2}
+\mu^3\arctan\left[
\frac{F}{\sqrt{\mu^2-F^2}}\right]
-F^3\ln\frac{\mu+\sqrt{\mu^2-F^2}}{F}\right]
\nonumber\\
& &\quad\quad\times
(\mu_u+\mu_d)\nonumber\\
&=&\frac{F}{2G}(\mu_u+\mu_d)\ , 
\label{6-8b}
\eeq
\esub
where the gap equation $\partial \Phi/\partial F
=\sum_{f=u,d}{\partial \Phi}/\partial {\wtilde F}_f +\partial (F^2/(2G))/\partial F=0$ 
is used from the second line to the third line in (\ref{6-8b}). 
It should be noted that if $F=0$, the spontaneous magnetization disappears because 
$F=0$ in Eq.(\ref{6-8b}). 

In Fig.\ref{fig:fig2}, 
the spontaneous magnetization is depicted as a function of the quark chemical potential. 
The spontaneous magnetization is shown in SI unit.
For $\mu<\mu_c\approx 0.407$ GeV, the magnetization does not occur because the spin polarization does not appear, $F=0$. 
At $\mu=\mu_c$, the spontaneous magnetization suddenly appears.

\setcounter{equation}{0}
\section{Magnetic field of hybrid compact star}

%
\begin{figure}[t]
\begin{center}
\includegraphics[height=5cm]{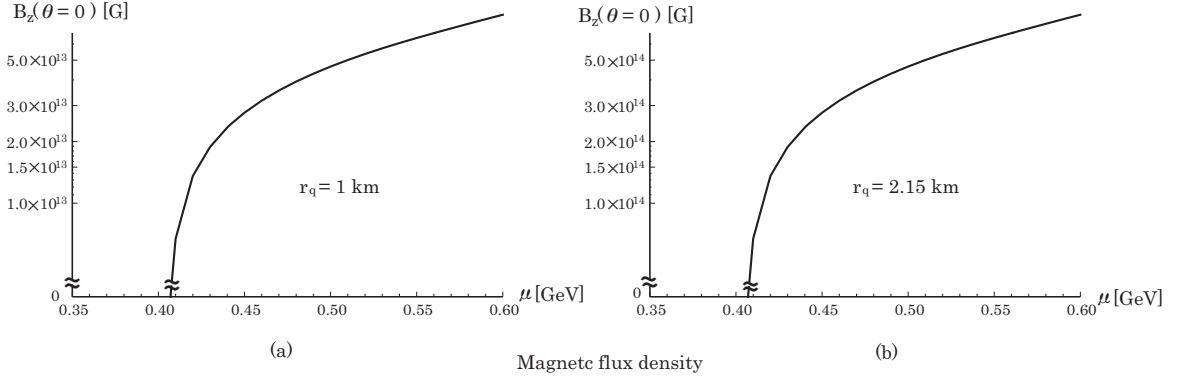}
\caption{The magnetic flux density is shown in a logarithmic scale as a function of the quark chemical potential $\mu$ 
in the case (a) $r_q=1$ km where the quark matter occupies 0.1 \% of the total volume of star 
($(4\pi r_q^3/3)/(4\pi R^3/3)=0.001)$
and (b) $r_q=2.15$ km where the quark matter 
occupies 1 \% of the total volume of star ($(4\pi r_q^3/3)/(4\pi R^3/3)=0.01)$, respectively. 
It should be noted that the SI unit, Tesla, is converted to Gauss, namely, 1 T (Tesla) $=10^4$ G (Gauss).
}
\label{fig:fig3}
\end{center}
\end{figure}
%

As seen in the previous section, for $\mu > \mu_c$, 
the spontaneous magnetization in the quark matter appears. 
Here, the spontaneous magnetization per unit volume ${\cal M}$ may be regarded as the magnetic dipole moment. 
Under this identification, we can calculate the strength of the magnetic filed yielded by the spontaneous magnetization.
As is well known in the classical electromagnetism, 
the magnetic field at the position ${\mib r}$, namely magnetic flux density ${\mib B}({\mib r})$, created by the magnetic dipole moment 
${\mib m}$ 
can be expressed as 
\beq\label{7-1}
{\mib B}=\frac{\mu_0}{4\pi}\left[-\frac{{\mib m}}{r^3}+\frac{3{\mib r}({\mib m}\cdot{\mib r})}{r^5}\right]\ , 
\eeq
where $\mu_0$ represents the vacuum permeability.
In our case, ${\mib m}=(0,0,{\cal M}\times V)$, where $V$ represents a volume because ${\cal M}$ is nothing but the magnetization per unit volume.

Here, let us consider the hybrid star with quark matter in the core of neutron star. 
Let us assume that the hybrid (neutron) star has radius $R=10$ km. 
If there exists quark matter in the inner core of star from the center to $r_q$ km, 
the strength of the magnetic flux density on the surface 
at the north or south pole of hybrid star is roughly estimated as 
\beq\label{7-2}
B_z=\frac{\mu_0}{4\pi}\left(-\frac{{\cal M}}{R^3}+\frac{3z^2{\cal M}}{R^5}\right)\times \frac{4}{3}\pi r_q^3=\mu_0\frac{2{\cal M}r_q^3}{3R^3}\quad [T]\ .
\eeq
Figure \ref{fig:fig3} (a) and (b) 
shows the magnetic flux density as a function of the quark chemical potential $\mu$ 
in the case (a) $r_q=1$ km and (b) $r_q=2.15$ km where quark matter 
occupies (a) 0.1 \% and (b) 1 \% of the total volume of star ($(4\pi r_q^3/3)/(4\pi R^3/3)= {\rm (a)}\ 0.001\ {\rm and\ (b)}\ 0.01)$, respectively. 
It should be noted that the SI unit, Tesla, is converted to Gauss, namely, 1 T (Tesla) $=10^4$ G (Gauss).
In the magnetar, the strength of the magnetic field at the surface of star is near $10^{15}$ G. 
Thus, in our calculation, if quark matter exists and the spin polarization occurs, 
a strong magnetic field about $10^{13}$ or $10^{14}$ Gauss may be created.

\section{Summary and concluding remarks}

It has been shown that spontaneous magnetization occurs due to the anomalous magnetic moments of 
quarks in the high-density quark matter under the tensor-type four-point interaction. 
In the Nambu-Jona-Lasinio model as an effective model of QCD, the tensor-type four-point 
interaction has been introduced. 
As for the tensor-type four-point interaction, this interaction term was also introduced to investigate the meson spectroscopy, 
especially, for vector and axial-vector mesons \cite{tensor}. 
Owing to this interaction, the spin polarized condensate appears for each quark flavor 
in the region of large quark chemical potential.  
It has been shown that the spin polarized condensate leads to spontaneous magnetization 
of quark matter due to the anomalous magnetic moments of quarks. 
Also, it has been pointed out that spontaneous magnetization 
does not occur if there exists no anomalous magnetic moments of quarks.

In this paper, furthermore, the implication to the strong magnetic field in compact stars 
such as the hybrid stars has been discussed. 
If there exists quark matter in the core of neutron stars and then the quark number density is 
rather high, spontaneous magnetization may occurs. 
If quark matter occupies a volume with 1\% of neutron star, 
the strength of magnetic field at the surface of neutron star is of order of $10^{14}$ Gauss, 
which is comparable to the strength of magnetic field in the so-called magnetar.

As indicated in this paper, the spin polarized condensate appears in the region with a 
large quark chemical potential. 
Thus, the chiral symmetry is broken in this model because the starting Lagrangian density is 
constructed with chiral symmetry. 
Under the magnetic field, the chiral symmetry is broken and it is shown that 
the chiral condensate grows at most linearly as a function of the magnetic field $B$ \cite{Kojo}. 
Further, in Ref.\citen{Ferrer}, the tensor-type four-point interaction was introduced in the 
NJL model in the case of one quark-flavor. 
In that paper, both the chiral condensate and spin polarized condensate are equally treated 
at finite temperature system. 
Thus, it may be an interesting problem that the coexistence of 
the chiral condensate and the spin polarized condensate is also considered at 
finite baryon density system.

As for the implication to compact stars, it has been shown that there is a possibility of the existence of the massive hybrid quark stars 
with two solar mass under the strong magnetic field \cite{Sotani}. 
Thus, the investigation of the equation of state of quark matter in the spin polarized phase revealing the spontaneous magnetization 
may be one of the interesting future problems.

\section*{Acknowledgment}

Two of the authors (Y.T. and M.Y.) would like to express their sincere thanks to\break
Professor J. da Provid\^encia and Professor C. Provid\^encia, two of co-authors of this paper, 
for their warm hospitality during their visit to Coimbra in spring of 2015. 
One of the authors (Y.T.) 
is partially supported by the Grants-in-Aid of the Scientific Research 
(No.26400277) from the Ministry of Education, Culture, Sports, Science and 
Technology in Japan.

\vspace{-0cm}





\begin{thebibliography}{99}


\bibitem{FH}
K. Fukushima and T. Hatsuda, Rep. Prog. Phys. {\bf 74}, 014001 (2011).

\bibitem{ARW}
M. Alford, K. Rajagopal and F. Wilczek, Nucl. Phys. B {\bf 537}, 443 (1999).

\bibitem{IB}
K. Iida and G. Baym, Phys. Rev. D {\bf 63}, 074018 (2001). 

\bibitem{CFL}
M. G. Alford, A. Schmitt, K. Rajagopal and T. Schafer, Rev. Mod. Phys. {\bf 80}, 1455 (2008) and references cited therein.

\bibitem{MaC}
L. McLerran and R. D. Pisarski, Nucl. Phys. A {\bf 796}, 83 (2007). 


\bibitem{NT}
E. Nakano and T. Tatsumi, Phys. Rev. D {\bf 71}, 114006 (2005). 


\bibitem{Blaschke}
T. Kl\"ahn, R. \L astowiecki and D. Blaschke, Phys. Rev. D {\bf 88}, 085001 (2013). 

\bibitem{magnetar1}
R. C. Duncan and C. Thompson, Astrophys. J. {\bf 392}, L9 (1992).

\bibitem{magnetar2}
C. Thompson and R. C. Duncan, Astrophys. J. {\bf 408}, 194 (1993), ibid. {\bf 473}, 322 (1996).


\bibitem{Tatsumi}
T. Tatsumi, Phys. Lett. B {\bf 489}, 280 (2000). 


\bibitem{NMT}
E. Nakano, T. Maruyama and T. Tatsumi, Phys. Rev. D {\bf 68}, 105001 (2003). 

\bibitem{TMN}
T. Tatsumi, T. Maruyama and E. Nakano, Prog. Theor. Phys. Suppl. No. 153, 190 (2004).



\bibitem{BJ}
H. Bohr, P. K. Panda, C. Provid\^encia and J. da Provid\^encia, Braz.\ J. Phys. {\bf 42}, 68 (2012).


\bibitem{arXiv}
H. Bohr, P. K. Panda, C. Provid\^encia and J. da Provid\^encia, 
Int. J. Mod. Phys. E {\bf 22}, 1350019 (2013).


\bibitem{oursPTP}
Y. Tsue, J. da Provid\^encia, C. Provid\^encis and M. Yamamura, 
Prog. Theor. Phys. {\bf 128}, 507 (2012).



\bibitem{oursPTEP1}
Y. Tsue, J. da Provid\^encia, C. Provid\^encis, M. Yamamura and H. Bhor, 
Prog. Theor. Exp. Phys. {\bf 2013}, 103D01 (2013).



\bibitem{oursPTEP2}
Y. Tsue, J. da Provid\^encia, C. Provid\^encis, M. Yamamura and H. Bhor, 
Prog. Theor. Exp. Phys. {\bf 2015}, 013D02 (2015).


\bibitem{NJL}
Y. Nambu and G. Jona-Lasinio, Phys. Rev. {\bf 122}, 345 (1961), Phys. Rev {\bf 124}, 246 (1961).

\bibitem{Buballa}
M. Buballa, Phys. Rep. {\bf 407}, 205 (2005). 


\bibitem{AMM}
R.G. Felipe, A. P. Mart\'inez, H. P. Rojas and M. Orsaria, 
Phys. Rev. C. {\bf 77}, 015807 (2008).

\bibitem{tensor}
M. Jaminon, E. Ruiz Arriola, Phys. Lett. B {\bf 443}, 33 (1998).


\bibitem{Kojo}
T. Kojo and N. Su, Phys. Lett. B {\bf 720}, 192 (2013). 

\bibitem{Ferrer}
E. J. Ferrer, V. de la Incera, I. Portillo and M. Quiroz, Phys. Rev. D {\bf 89}, 085034 (2014). 



\bibitem{Sotani}
H. Sotani and T. Tatsumi, Mon, Not. R. Astron. Soc. {\bf 447}, 3155 (2015). 

\end{thebibliography}
%


\end{document}